\documentstyle [12pt,twoside]{article}
  
\begin{document}
\vspace*{-2.cm} 
\centerline{\large{\bf Nonstationary Stochastic Resonance
in a Single Neuron-Like System}\normalsize} 
\vspace*{0.5cm}
\centerline{\bf Redouane Fakir}
\vspace*{.5cm}
\centerline{\em Peter Wall Institute for Advanced Studies}
\centerline{\em Cosmology Group, Department of Physics \& Astronomy}
\centerline{\em Department of Interdisciplinary Studies}  
\centerline{\em University of British Columbia, 
6224 Agriculture Road, Vancouver, B. C. V6T 1Z1, CANADA}
\vspace*{1.0cm}
{\bf
Stochastic resonance holds much promise for the detection
of weak signals in the presence of relatively loud noise.
Following the discovery of nondynamical and of aperiodic
stochastic resonance, it was recently shown that the phenomenon
can manifest itself even in the presence of 
nonstationary signals. This was found in a composite system of 
differentiated trigger
mechanisms mounted in parallel, which suggests that it could
be realized in some elementary neural networks or nonlinear electronic
circuits. Here, we find
that even an individual trigger system may be able to detect
weak nonstationary signals using stochastic resonance. The very
simple modification to the trigger mechanism that makes this possible
is reminiscent of some aspects of actual neuron physics. Stochastic
resonance may thus become relevant to more types of
biological or electronic systems injected with an ever broader class
of realistic signals.
}

\clearpage
One of the remarkable aspects of stochastic resonance$^{1-10}$
is the possibility of enabling the detection of a
weak signal by adding noise to the input.
An important recent development in this field - the discovery of 
nondynamical stochastic resonance$^{11-13}$ - is that the phenomenon can
occur independently of any {\em details} of nonlinear dynamics in the
system, although the nonlinearity itself is essential for the phenomenon
to occur. Thus, stochastic resonance was shown to arise in extremely
simple trigger systems.
Equally
important was the discovery of aperiodic stochastic resonance$^{14-23}$,
that is, the realization that the signals made detectable by the addition
of noise need not be periodic. This opened the door for investigating
the occurrence of the phenomenon under a broader set of realistic conditions.
More recently still, a study originally motivated by new prospects in
gravitational wave detection$^{24-26}$ showed
that stochastic resonance can manifest itself not only when the
signal is aperiodic, but also when it is markedly nonstationary$^{27}$.
This could potentially extend the relevance of stochastic resonance
to a larger class of biological and electronic applications.
The system that was shown to exhibit nonstationary stochastic resonance
was a simple, nondynamical, multi-level trigger, more precisely, a summing
network of differentiated single-threshold systems. Here, we report
that nonstationary stochastic resonance can also manifest itself 
in a system as elementary as an individual, single-level trigger, provided
a very simple modification is applied to the trigger mechanism. 
This modified single-threshold system is reminiscent of certain
aspects of neuronal biophysics, and we shall briefly allude to that
possible connection further below.

Consider then the single-threshold trigger mechanism that is at work
in Figs.1,2. Starting with Fig.1a, it shows the input consisting of a sub-threshold deterministic signal (see Fig.1b)
buried in loud (i.e. above-threshold) random noise. The latter is
taken to be a low-pass filtered, zero-mean Gaussian white noise.
The total input frequently exceeds the threshold, resulting in the
firings of Figs.2, although the deterministic signal itself
never exceeds the threshold, and hence would not be detectable
if it were the only input. 

The injected deterministic signal is markedly nonstationary, while always
remaining sub-threshold. Hence, no response would result in Figs.2
in the absence of noise. In the presence of noise, not only one
does obtain a response, but the hidden deterministic signal can be
easily detected in that response, as can be seen most clearly from
Fig.2c. It is the strong correlation of the noise-induced response
with the sub-threshold signal (Figs.3,4) that permits one to say that the 
addition of noise has allowed the detection of an otherwise
undetectable signal. Note that, if the signal has more high-frequency
structure than shown here, that could be dealt with by first applying
the techniques of aperiodic stochastic resonance$^{14-23}$ to 
the ``straightened out''
(low-frequency filtered) version of the signal. In the present letter,
we focus on the nonstationary aspect of the signal, which is essentially
a low-frequency characteristic.

The trigger system, in its simplest form, is sensitive only to 
{\em whether} (not to {\em by how much}) the threshold is exceeded.
Hence, one does not expect that a single-trigger system would  help 
efficiently with the detection of a strongly nonstationary signal.
This is indeed confirmed in Fig.3a, where
the response of the unmodified trigger system is shown to have little visible
structure and to correlate poorly with the hidden deterministic signal.

This eventual inability of the simplest single-trigger system to help
efficiently with the detection of a strongly nonstationary signal
can sometimes be remedied if several such triggers are available and
if their outputs can be summed$^{27}$. That not withstanding,
it is shown in Figs.2b,c that even one individual trigger can help achieve
detection, provided that either one of the two following straightforward modifications can be made: 

1) The system fires only every time interval
equal to $B$, and the height of the pulse fired is proportional to the
number of times the threshold has been crossed during $B$. Equivalently,
the system outputs after every time interval $B$ the average of the
``blind'' response of Fig.2a. This situation can easily arise when, e.g.,
the response of a system is (by choice or by constraint) slower than the 
input sampling rate. Indeed, when the system is one under biological
or electronic control, efficiency dictates that the output firing rate
{\em should be} slower than the input sampling rate.
Fig.2b shows that this modification can improve even visually the correlation with the hidden deterministic signal, which improvement is confirmed more quantitatively in Fig.3.

2) The internal dynamics of the system effects the response in a way
that can be modeled by a convolution window running through the pulsed
``blind'' response (i.e. through Fig.2a.) This is indeed expected to be 
the case if the system simulated is, e.g., a cortical or a sensory 
neuron$^{16,28-32}$. 
The minimal lapse between two neural firings can be as small as a few milliseconds, but there is an additional, longer time scale effecting 
the response, a time scale that can vary from about $10$ milliseconds for certain cortical neurons, to about $100$ milliseconds for sensory neurons. 
This effective integration time reflects the characteristics
of the flow of ionic transmitter substances through dendritic synapses and  the 
associated growth of the polarizing potentials involved in the neural
firings. In Fig.2c, the system has been modified to reflect in
the simplest way possible the basic aspects of neuron physics just mentioned: 
An effective running window, taken here to be a Gaussian of width $B$, is applied to the raw pulse train of Fig.2a (i.e., the actual response is the
convolution of the blind response of Fig.2a with a Gaussian of width $B$.)
For most values of $B$, the response of this modified trigger is extremely well correlated to the hidden deterministic signal (see Fig.3). Simulations show
that the results remain virtually unchanged for most reasonable choices of 
a running window. This implies that if a given type of neuron is actually found
to display the phenomenon suggested by these simulations, the same could
be suspected to hold for most other types of neurons.
 
To summarize, it can be seen from Fig.2b and Fig.2c that elementary,
experimentally motivated modifications of the single-trigger system 
can produce a dramatic increase in the efficiency of signal detection
through stochastic resonance.
The implication is that even one individual neuron, or an analogous 
nonlinear electronic device, could help achieve the detection of nonstationary, sub-threshold signals in a noisy environment.
 
Coming in the wake of the remarkable leaps in the field brought
about by nondynamic and aperiodic stochastic resonance, and following the 
recently suggested generalization to markedly nonstationary cases, the possibility seems to become ever more real that some of the simplest 
systems conceivable may be able to detect weak signals of an almost arbitrary nature. 

\vspace*{2.cm}
\centerline{\bf Acknowledgements}
\vspace*{0.5cm}
I am grateful to L.M. Ward for helping me become better acquainted with
neural processes. I am also grateful to W.G. Unruh for being the first
to bring the phenomenon of stochastic resonance to my attention, and to
B. Bergersen for helping me familiarize myself with previous literature
on stochastic resonance. I have also benefited from  
extensive logistical support by the General Relativity
\& Cosmology Group in the Department of Physics, University of
British Columbia, by the Department of Interdisciplinary Studies
at UBC and by T.E. Vassar during the preparation of this paper. 
This work was supported by the PWIAS of UBC and NSERC of Canada.

\clearpage
\hspace*{4.cm} {\bf\large REFERENCES}
\newline\newline\newline
1. Benzi R., Sutera S. \& Vulpiani A, {\em J. Phys.} A{\bf 14},
L453 (1981).
\newline\newline
2. Nicolis C., {\em Tellus} {\bf 34}, 1 (1982).
\newline\newline
3. Benzi R., Parisi G., Sutera A \& Vulpiani A., {\em Tellus} {\bf 34},
10 (1982).
\newline\newline
4. Wiesenfeld K. \& Moss F., {\em Nature} {\bf 373}, 33 (1995).
\newline\newline
5. Nicolis C., {\em J. Stat. Phys.} {\bf 70}, 3 (1993).
\newline\newline
6. Longtin A., Bulsara A. \& Moss F, {\em Phys. Rev. Lett.}{\bf 67},
656 (1991).
\newline\newline
7. Douglass J.K., Wilkens L., Pantazelou E. \& Moss F.,
{\em Nature} {\bf 365}, 337 (1993).
\newline\newline
8. Bezroukov S.M. \& Vodyanoy I., {\em Nature} {\bf 378}, 362 (1995).
\newline\newline
9. Fauve S. \& Heslot F., {\em Phys. Lett.} A{\bf 97}, 5 (1983).
\newline\newline
10. Mantegna R.N. \& Spagnolo B., {\em Phys. Rev.} E{\bf 49} R1792 (1994).
\newline\newline
11. Jung P., {\em Phys. Rev.} E{\bf 50}, 2513 (1994);
{\em Phys. Lett.} A{\bf 207}, 93 (1995).
\newline\newline
12. Wiesenfeld K., Pierson D., Pantazelou E., Dames C. \& Moss F.,
{\em Phys. Rev. Lett.} {\bf 72}, 2125 (1994).
\newline\newline
13. Gingl Z., Kiss L.B. \& Moss F., {\em Europhys. Lett.} {\bf 29},
191 (1995).
\newline\newline
14. Collins J.J., Chow C.C. \& Imhoff T.T., {\em Phys. Rev.}
E{\bf 52}, R3321 (1995).
\newline\newline
15. Collins J.J., Chow C.C., Capela A.C. \& Imhoff T.T., {\em Phys. Rev.}
E{\bf 54}, 5575 (1996).
\newline\newline
16. Collins J.J., Imhoff T.T. \& Grigg P., {\em J. of Neurophysiology}
{\bf 76}, 642 (1996).
\newline\newline
17. Collins J.J., Chow C.C. \& Imhoff T.T., {\em Nature} {\bf 376}, 236 (1995).
\newline\newline
18. Levin J.E. \& Miller J.P., {\em Nature} {\bf 380}, 165 (1996).
\newline\newline
19. Heneghan C., Chow C.C., Collins J.J., Imhoff T.T.,
Lowen S.B. \& Teich M.C., {\em Phys. Rev.} E{\bf 54}, R2228 (1996).
\newline\newline
20. Gailey P.C., Neiman A., Collins J.J. \& Moss F., 
{\em Phys. Rev. Lett.} {\bf 79}, 4701 (1997).
\newline\newline
21. Neiman A., Schimansky-Geier L. \& Moss F., {\em Phys. Rev.} E{\bf 56},
9 (1997).
\newline\newline
22. Pantazelou E., Moss F. \& Chialvo D., in {\em Noise in Physical Systems
and 1/f Fluctuations}, (eds Handel P.H. and Chung A.L.) 549-552 
(American Institute of Physics Press, New York, 1993.)
\newline\newline
23. Moss F. \& Pei X., {\em Nature} {\bf 376}, 221 (1995).
\newline\newline 
24. Fakir R., {\em Int. J. Mod. Phys.} D{\bf 6}, 49 (1997);
{\em Phys. Rev.} D {\bf 50}, 3795 (1994); {\em Astrophys. J.} {\bf 426},
74 (1994).
\newline\newline
25. Unruh W.G., private communication (1995).
\newline\newline
26. Fakir R., {\em ``A macroscopic gravity wave effect;''} Los Alamos Physics Archives astro-ph/9601127.
\newline\newline
27. Fakir R., {\em ``Nonstationary Stochastic Resonance;''} to appear in
{\em Phys. Rev.} E (1998).
\newline\newline
28. Luce R.D. \& Green D.M., {\em Psychological Review} {\bf 79}, 14 (1972).
\newline\newline
29. Ward L.M., {\em Perception \& Psychophysics} {\bf 50}, 117-128 (1991).
\newline\newline
30. Hecht S., Schlaer S. \& Pirenne M.H., {\em J. of General Physiology}
{\bf 25}, 819-840 (1942).
\newline\newline
31. Seamans J.K., Gorelova, N. \& Yang, C.R. {\em J. Neuroscience} {\bf 17},
5936-5948 (1997).
\newline\newline
32. K\"{o}nig P., Engel A.K. \& Singer W. {\em Trends in Neuroscience} {\bf 19},
130-137 (1996).
 
\clearpage
\centerline{\large\bf Figure captions}
\vspace{4.cm}
\centerline{\bf Figure 1}
\vspace{2.cm}
{\em
a) The input consists of a sub-threshold deterministic signal (Fig.1b)
buried in random noise. Here as throughout the simulations, the total integration time is normalized to one.

b) The deterministic signal is markedly nonstationary, while always
remaining sub-threshold. No response would result in Figs.2
without the presence of noise.
}
\clearpage
\vspace{4.cm}
\centerline{\bf Figure 2}
\vspace{2.cm}
{\em
a) Response of the unmodified trigger system, showing little visible
structure and correlating poorly to the hidden deterministic signal.
 
b) This is the response from an individual trigger system that averages its
raw pulse train (the firings of Fig.2a) every time interval $B$.
Equivalently, this would be the response
from a system that, after every time interval $B$, fires a pulse with a height
proportional to the number of times during $B$ that the threshold has been exceeded.
 
c) The single-trigger system has been modified here in keeping with
some basic facts about internal neuron physics (see text).  
A Gaussian running window of width $B$ is applied to 
the raw pulse train of Fig.2a. For most values of $B$, the response 
is strongly correlated to the hidden deterministic signal (see Fig.1b).
}
\clearpage
\vspace{4.cm}
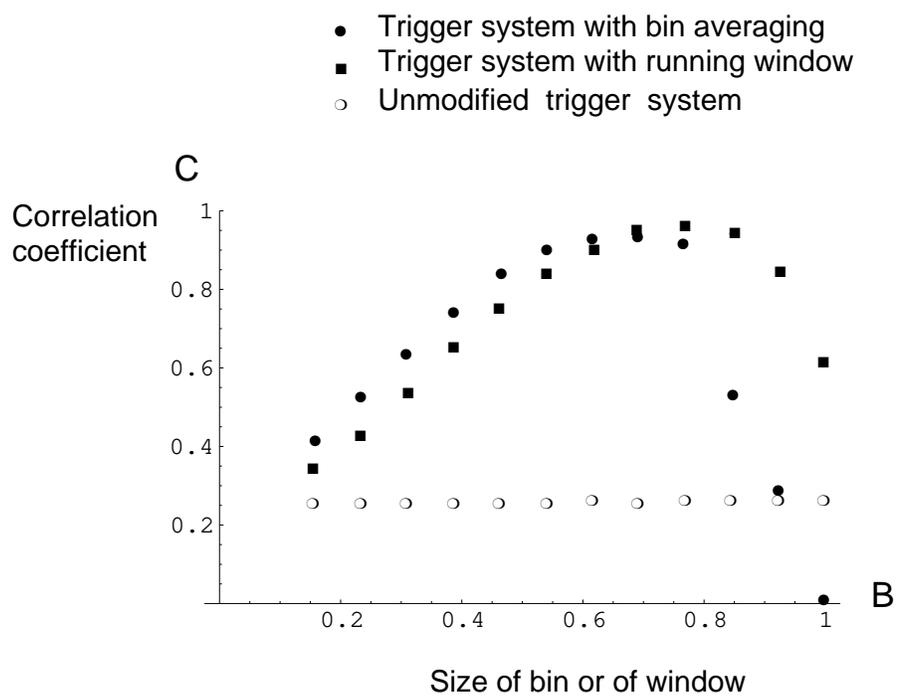
\centerline{\bf Figure 3}
\vspace{2.cm}
{\em
This shows the correlation coefficient, which is the 
value of the normalized correlation function (between the response and the
signal) at zero-lag, as a function of $B$, the size of the averaging bin
or of the running window. The very high correlations achieved for certain
values of $B$ confirm the visual detection of the deterministic signal
(Fig.1b) in the responses of Figs.2b,c. 
}
\clearpage
\vspace{4.cm}
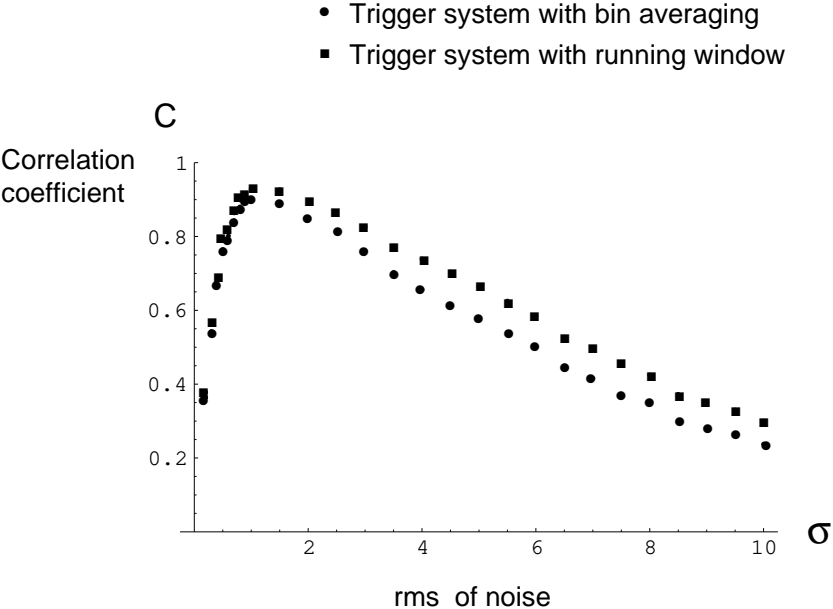
\centerline{\bf Figure 4}
\vspace{2.cm}
{\em
The correlation coefficient (see caption of Fig.3) as a function of $\sigma$,
the rms of the noise, which is divided here by threshold height of Figs.1a,b. The system clearly displays stochastic resonance. 
}
 
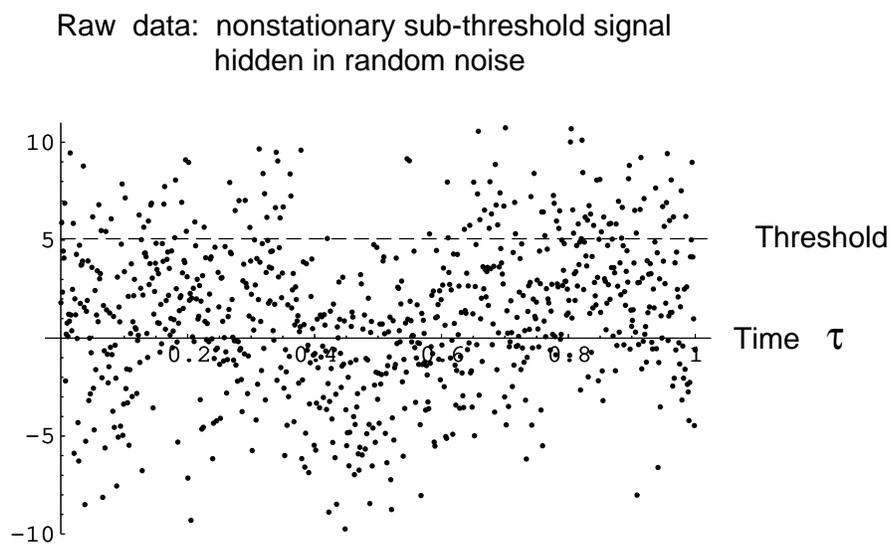
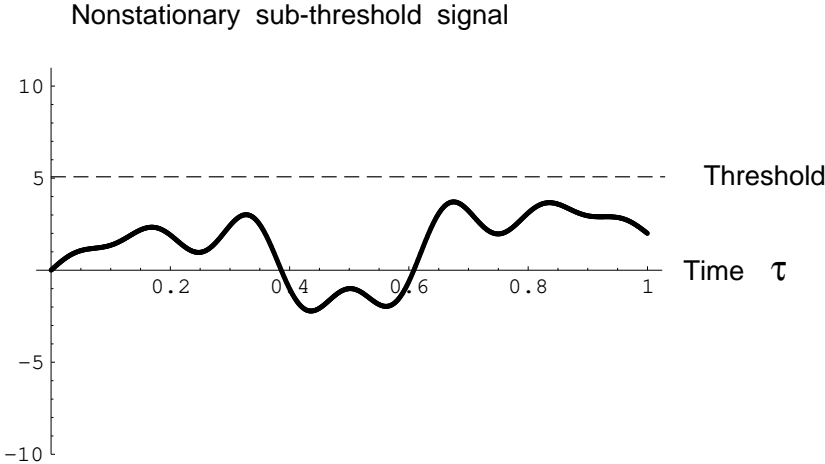
\end{document}